\documentclass[lettersize,journal]{IEEEtran}

\usepackage{amsmath,amsfonts}
\usepackage{algorithmic}
\usepackage{algorithm}
\usepackage{array}

\usepackage[caption=false,font=normalsize,labelfont=sf,textfont=sf]{subfig}

\usepackage{textcomp}
\usepackage{stfloats}
\usepackage{url}
\usepackage{verbatim}
\usepackage{graphicx}
\usepackage{cite}
\usepackage{bm}
\usepackage{amsmath}
\usepackage[flushleft]{threeparttable}
\usepackage{booktabs}
\usepackage[T1]{fontenc}
\usepackage{newtxmath}

\begin{document}

\title{Risk-Aware Safe Throughput Forecasting for Starlink Networks}
\author{Hongjun Xie, Chao Zhang, Pengcheng Luo, Zenghui Zhang, Genke Yang, Xiaojuan Zhang, \IEEEmembership{Senior Member, IEEE}, \\and Boon-Hee Soong, \IEEEmembership{Life Senior Member, IEEE}\thanks{This work was supported by the National Major Science and Technology Project for Intelligent Manufacturing Systems and Robotics of China under Grant 2025ZD1602400. \emph{(Corresponding author: Pengcheng Luo.)}}
\thanks{Hongjun Xie,  Pengcheng Luo, Zenghui Zhang, and Genke Yang are with Ningbo Artificial Intelligence Institute, Shanghai Jiao Tong University, Ningbo 315000, China, and also with the School of Automation and Intelligent Sensing, Shanghai Jiao Tong University, Shanghai 200240, China, and the Key Laboratory of System Control and Information Processing, Ministry of Education of China, Shanghai 200240, China (e-mail: xiehongjun@sjtu.edu.cn, luopeng69131@sjtu.edu.cn, zenghui.zhang@sjtu.edu.cn, gkyang@sjtu.edu.cn).}
\thanks{Chao Zhang is with the National Key Laboratory of Wireless Communications, University of Electronic Science and Technology of China, Chengdu 611731, China (e-mail: zhang\_chao@std.uestc.edu.cn).}
\thanks{Xiaojuan Zhang is with the Institute for Infocomm Research, A*STAR, Singapore 138632 (e-mail: xiaojuanzhang@ieee.org).}
\thanks{Boon-Hee Soong is with the School of Electrical and Electronic Engineering, Nanyang Technological University, Singapore 639798 (e-mail: ebhsoong@ntu.edu.sg).}
}

\maketitle

\begin{abstract}
As a representative low Earth orbit (LEO) broadband system, Starlink exhibits highly variable access throughput, making short-term forecasting essential for network resource management. Existing forecasting methods mainly optimize symmetric point-prediction metrics such as MAE and RMSE, but they do not explicitly control the asymmetric risk of overestimating future throughput, which can cause over-admission, bandwidth overbooking, and service violations. This paper formulates Starlink throughput prediction as a risk-budgeted safe forecasting problem, where the predictor must satisfy a prescribed overestimation budget while maintaining competitive accuracy. We propose Budget-Guided Coarse-to-Fine Quantile Selection (BG-CFQS), a data-driven framework that trains a family of lower-quantile predictors, locates the quantile boundary satisfying the risk budget, and refines the boundary region to select the most accurate feasible predictor. Experiments on three real-world Starlink throughput datasets show that BG-CFQS satisfies the risk budget on all datasets and achieves the lowest average MAE, mean positive error, and tail positive error among budget-feasible methods. In high-risk and severe-risk low-throughput regimes, BG-CFQS reduces harmful positive errors by 11.0\% and 12.6\%, respectively. An admission-control evaluation further shows that the proposed safe forecasts reduce dropped sessions, demonstrating that risk-aware forecasting can translate prediction safety into application-level benefits.

\end{abstract}


\section{Introduction}
\subsection{Background and Motivation}
\IEEEPARstart{S}{tarlink} has become the most widely deployed low Earth orbit (LEO) broadband network and an important component of next-generation Internet infrastructure. By leveraging a large constellation of low-altitude satellites, Starlink can provide broadband access to rural areas, oceans, airborne platforms, disaster recovery scenarios, and other regions where terrestrial connectivity is unavailable or economically difficult to deploy~\cite{zheng2023sdn}. Compared with traditional geostationary satellite systems, Starlink-like LEO broadband networks provide lower propagation latency and higher access capacity, making them attractive for latency-sensitive and bandwidth-intensive applications~\cite{li2024dynamic}.

Despite these advantages, Starlink access links exhibit substantial throughput variability. The serving satellite, elevation angle, propagation condition, obstruction status, gateway association, user mobility, and traffic load can all change over time~\cite{gao2024semantic}. Consequently, the throughput observed by Starlink users may fluctuate significantly even over short intervals. Short-term throughput forecasting is therefore important for bandwidth reservation, admission control, adaptive bitrate streaming, congestion control, edge offloading, and application-level quality-of-service management~\cite{lai2023starfront}.

\begin{figure*}[!t]
  \centering
  \includegraphics[width=\linewidth]{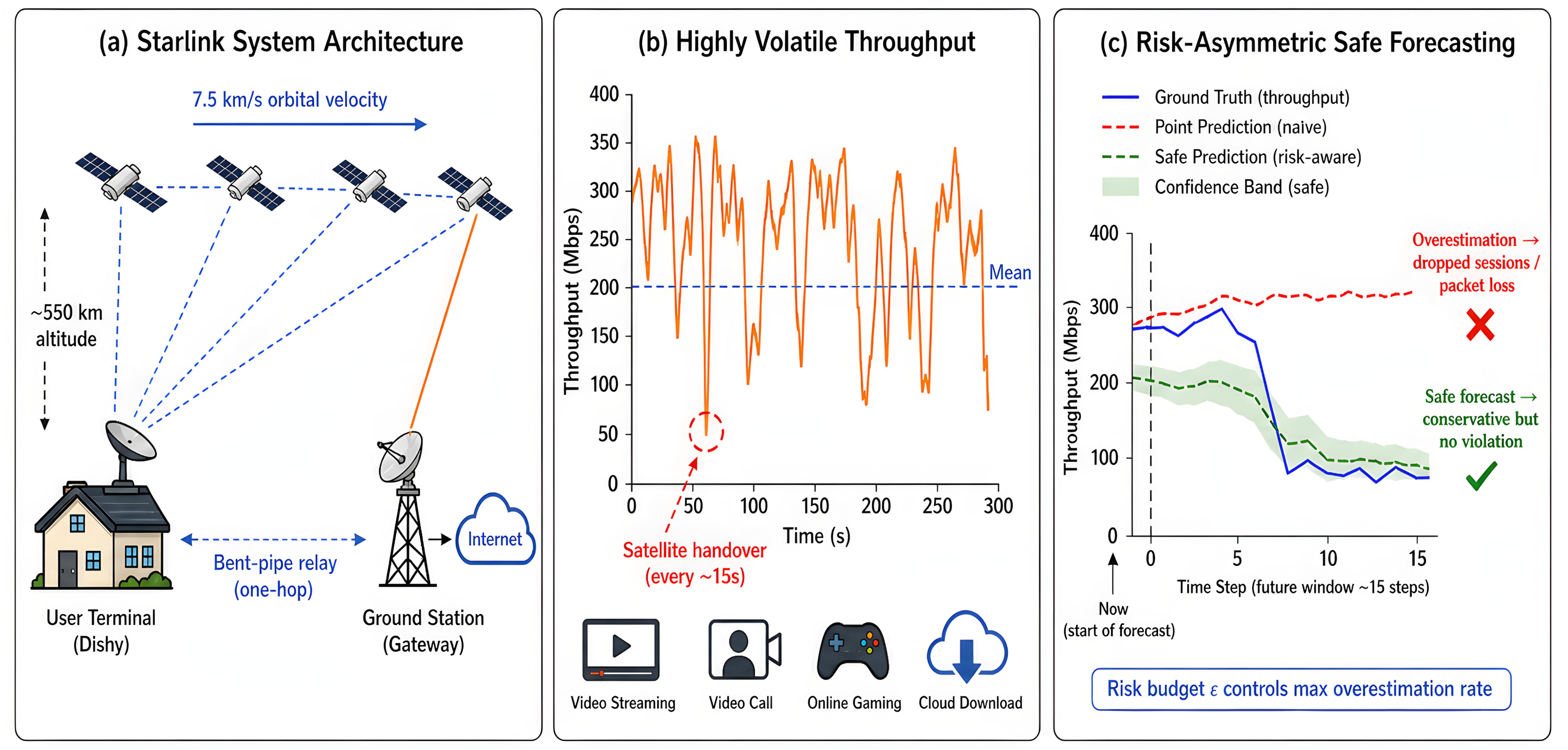}
  \caption{Overview of risk-aware safe throughput forecasting for Starlink networks.
  (a) Starlink bent-pipe architecture: the user terminal (dishy) connects via LEO satellites to ground stations for Internet access. Satellites orbit at $\sim$7.5 km/s, causing frequent handovers.
  (b) Throughput measured at the user terminal fluctuates significantly over short time scales due to satellite handovers, diurnal load patterns, and weather, making accurate short-term prediction essential.
  (c) Prediction errors are asymmetric: overestimating future throughput causes session drops and service violations, while a safe forecast avoids such harmful decisions at a modest cost in utilization. This work focuses on controlling the overestimation risk under a prescribed budget.}
  \label{fig:system-overview}
\end{figure*}

Most existing throughput prediction methods are formulated as point forecasting problems, where the objective is to minimize symmetric accuracy metrics such as mean absolute error (MAE) or root mean square error (RMSE)~\cite{schmid2021survey,mirza2010machine,narayanan2020lumos5g,mei2022realtime,minovski2021throughput}. Recent Starlink-oriented predictors, including T3P~\cite{tiwari2023t3p} and StarNet~\cite{liu2025starnet}, improve prediction accuracy by incorporating terminal telemetry, satellite-domain features, and handover-aware temporal patterns. These advances, however, still evaluate forecasts primarily through point accuracy, while downstream network decisions often incur asymmetric error costs. Underestimating future throughput mainly leads to conservative resource usage, whereas overestimating throughput can cause over-admission, excessive bitrate selection, queue buildup, packet loss, and service-level agreement violations. This asymmetry is especially relevant to Starlink networks, where LEO mobility, handovers, blockage, and scheduling changes can quickly reduce available bandwidth. Figure~\ref{fig:system-overview} illustrates the Starlink access architecture, throughput volatility, and the asymmetric consequence of unsafe overestimation.

This gap motivates a shift from conventional accuracy-oriented Starlink throughput prediction to \emph{risk-aware safe throughput forecasting}. Instead of only minimizing average prediction error, the forecast must remain usable by downstream decision modules under a prescribed overestimation risk budget. This requirement raises a calibration challenge: overly conservative predictions waste usable bandwidth, whereas insufficiently conservative predictions remain unsafe. Moreover, the appropriate conservative level may vary across datasets, network regimes, and application risk requirements. A practical safe forecasting method should therefore select the prediction level from data while explicitly respecting the risk budget.

To this end, we propose \emph{Budget-Guided Coarse-to-Fine Quantile Selection} (BG-CFQS), a risk-calibrated framework for safe throughput forecasting in Starlink networks. BG-CFQS trains a family of lower-quantile throughput predictors, locates the boundary quantile whose overestimation risk is close to the prescribed budget, and refines the search around this boundary to select the most accurate feasible quantile. By avoiding a manually fixed conservative level, BG-CFQS provides an interpretable mechanism for balancing prediction accuracy and safety.

\subsection{Contributions}
The main contributions of this paper are summarized as follows:
\begin{itemize}
\item We formulate Starlink throughput prediction as a risk-budget constrained safe forecasting problem. Unlike conventional point forecasting that optimizes symmetric accuracy metrics, the proposed formulation explicitly constrains overestimation risk and evaluates both the frequency and severity of harmful positive errors.

\item We propose BG-CFQS, a budget-guided coarse-to-fine quantile selection framework for safe throughput forecasting. BG-CFQS trains a family of lower-quantile predictors, locates the data-dependent quantile boundary satisfying the overestimation budget, and refines the boundary region to select an accurate feasible predictor without manually fixing a conservative quantile level.

\item We develop a safety-oriented evaluation protocol for LEO satellite throughput forecasting. Beyond MAE and RMSE, the evaluation covers risk-budget feasibility, positive-error severity, risk-accuracy frontiers, high-risk and severe-risk low-throughput regimes, advanced time-series forecasting baselines, and an admission-control application scenario. Experiments on three real-world Starlink datasets show that BG-CFQS reduces harmful overestimation while preserving competitive prediction accuracy and improving downstream decision reliability.
\end{itemize}

\subsection{Organization}
The remainder of this paper is organized as follows. Section II reviews related work on satellite network measurement, throughput prediction, and risk-aware forecasting. Section III formulates the safe throughput forecasting problem. Section IV presents the proposed BG-CFQS framework. Section V describes the experimental setting and reports the experimental results. Finally, Section VI concludes the paper.

\section{Related Works}

We review related work along three lines. First, LEO satellite network measurement studies characterize the volatility of Starlink throughput and motivate the need for short-term forecasting. Second, network throughput prediction studies develop statistical, machine-learning, and deep-learning predictors, but most focus on point accuracy. Third, risk-aware prediction and safe decision-making methods provide tools for handling asymmetric error costs, yet existing networking applications rarely formulate throughput forecasting under an explicit one-sided overestimation budget. These three lines contextualize the need for risk-budgeted safe throughput forecasting.

\subsection{LEO Satellite Network Measurement}

The rapid deployment of LEO satellite mega-constellations has motivated extensive measurement studies of Starlink performance. Early studies characterized Starlink throughput, latency, and packet loss using active probing and browser-side measurements~\cite{michel2022first,kassem2022browser}. Subsequent end-user and multi-terminal measurements further showed that Starlink performance varies with satellite orbital patterns, user location, time of day, and application workload~\cite{ma2023network,mohan2024multifaceted}. These studies consistently indicate that LEO broadband links are more dynamic than conventional terrestrial broadband.

Recent work has further identified the sources of this variability. Garcia et al.~\cite{garcia2023multi} analyzed throughput dynamics across multiple timescales, from seconds-level handovers to hours-level diurnal patterns. Liu et al.~\cite{liu2025starnet} conducted a large-scale Starlink measurement campaign across three countries and isolated key drivers such as 15-second satellite handovers, load balancing, and weather-induced degradation. Tanveer et al.~\cite{tanveer2023making} reverse-engineered Starlink scheduling behavior by correlating obstruction maps with satellite orbital data, while Tiwari et al.~\cite{tiwari2023t3p} built LEOScope and analyzed latency inflation caused by handovers and bent-pipe routing. Laniewski et al.~\cite{laniewski2024wetlinks} released WetLinks to connect Starlink performance with longitudinal weather measurements.

In parallel, simulation and emulation platforms have improved the reproducibility of LEO network research. StarryNet~\cite{lai2023starrynet}, Hypatia~\cite{kassing2020exploring}, L2D2~\cite{vasisht2021l2d2}, and recent open measurement tools~\cite{izhikevich2024democratizing} provide ways to study constellation behavior, routing, and integrated space-terrestrial networks. Overall, these measurements and platform studies establish the volatility of LEO network performance. However, they mainly characterize the network or provide evaluation infrastructure; they do not address how throughput forecasts should be risk-calibrated when overestimation can lead to unsafe resource allocation.

\subsection{Network Throughput Prediction}

Throughput prediction has been extensively studied in terrestrial networks. Schmid et al.~\cite{schmid2021survey} surveyed client throughput prediction algorithms across wired and wireless networks, including statistical models, classical machine learning, and deep learning. Representative works use support vector regression for TCP throughput prediction~\cite{mirza2010machine}, spatial-temporal modeling for mmWave 5G throughput~\cite{narayanan2020lumos5g}, LSTM-based mobile bandwidth and handoff prediction~\cite{mei2022realtime}, and classical ML models such as random forests and gradient boosting for LTE/5G downlink throughput~\cite{minovski2021throughput,al2024machine}. These methods generally treat throughput prediction as a point forecasting problem and optimize symmetric metrics such as MAE, RMSE, or MAPE.

LEO satellite networks introduce additional dynamics that have motivated satellite-specific predictors. StarNet~\cite{liu2025starnet} incorporates satellite-domain knowledge through periodical embedding for the 15-second handover cycle and a 2D-to-3D projection algorithm for identifying serving satellites from orbital data and obstruction maps. T3P~\cite{tiwari2023t3p} integrates Starlink terminal telemetry, satellite position data, and historical traces using XGBoost- and LSTM-based predictors. In addition, general-purpose time-series forecasting architectures have been applied as strong backbones for throughput traces. DLinear~\cite{zeng2023dlinear}, PatchTST~\cite{nie2022time}, TimesNet~\cite{wu2022timesnet}, and iTransformer~\cite{liu2023itransformer} provide diverse model designs for capturing temporal patterns.

These predictors improve point accuracy by introducing better features, architectures, or satellite-domain priors. However, their objective remains accuracy-oriented: overestimation and underestimation are penalized symmetrically during training and evaluation. Our work is orthogonal to backbone design. Rather than proposing another point predictor, we study how throughput forecasts should be selected and evaluated under an explicit one-sided overestimation-risk budget.

\subsection{Risk-Aware Prediction and Safe Decision-Making}

The asymmetric cost of prediction errors has been recognized in several networking decisions. In adaptive bitrate streaming, MPC~\cite{yin2015control} and Pensieve~\cite{mao2017neural} use throughput estimates to guide bitrate selection, and later work integrates LEO-specific predictors into robust ABR controllers~\cite{tiwari2023t3p,liu2025starnet}. These systems show that prediction quality directly affects user experience and application utility. Nevertheless, they usually consume forecasts produced by point predictors or rely on controller-level robustness, rather than imposing an explicit risk constraint on the forecast itself.

Statistical tools for asymmetric prediction have also been widely studied. Quantile regression~\cite{koenker2001quantile} trains predictors at specified conditional quantiles using the pinball loss, enabling conservative lower-quantile forecasts. Probabilistic forecasting, prediction intervals, and conformal prediction can quantify uncertainty and provide coverage guarantees. However, in many applications these methods focus on two-sided uncertainty intervals or coverage calibration. They do not directly answer which single throughput value should be exposed to a network decision module when the application specifies a one-sided budget on overestimation risk.

Our work bridges this gap by formulating \emph{risk-budgeted safe forecasting} for Starlink throughput prediction. The goal is to output a single safe throughput forecast that maximizes accuracy while satisfying a prescribed overestimation-rate constraint. Unlike post-hoc scaling of point predictions, BG-CFQS learns a family of lower-quantile predictors and selects the most accurate budget-feasible quantile from data. This connects risk-aware statistical prediction with the practical requirement of safe bandwidth estimation in volatile LEO satellite networks.

\section{Problem Formulation}

\begin{figure*}[!t]
  \centering
  \includegraphics[width=\linewidth]{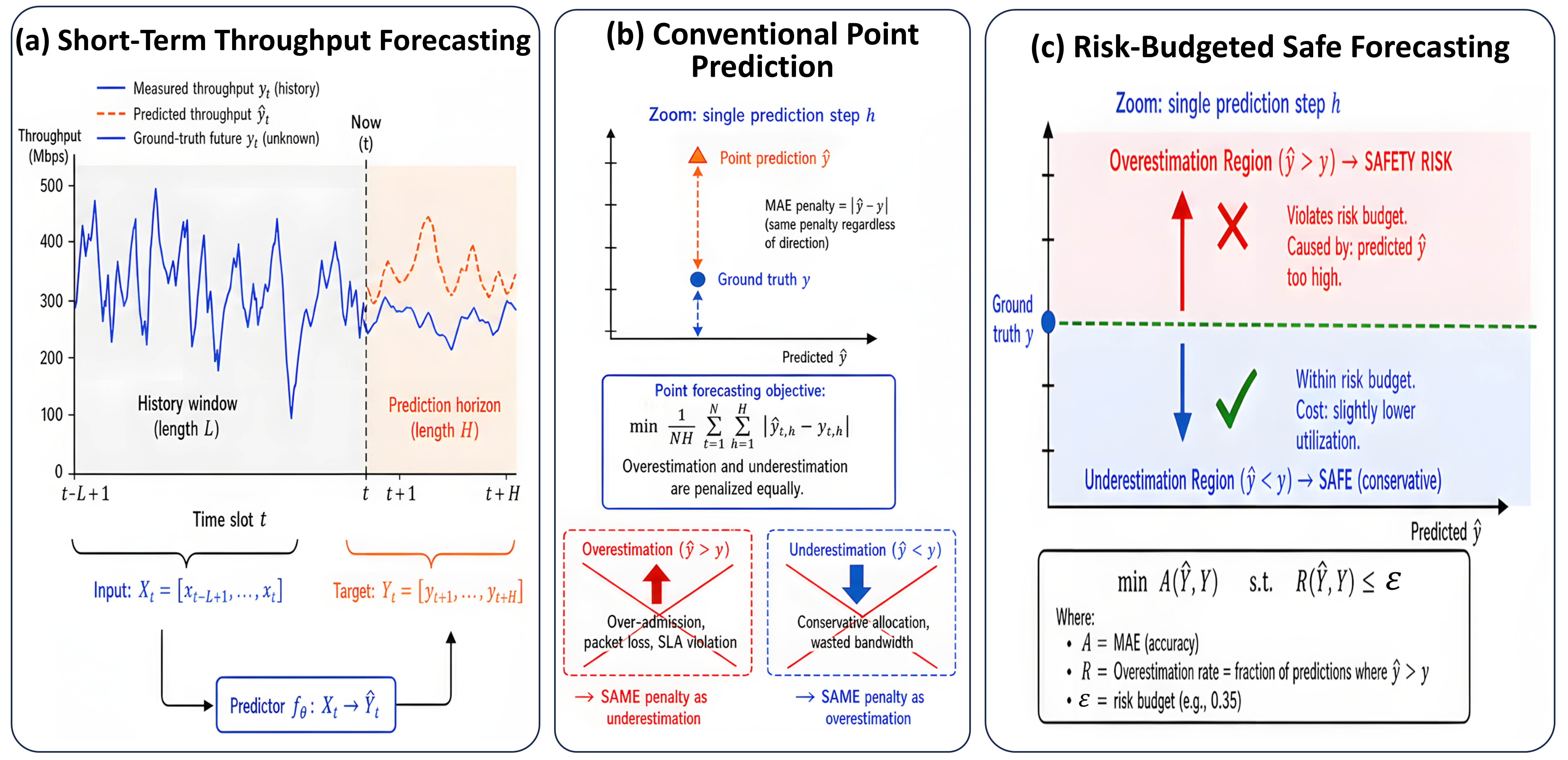}
  \caption{Overview of the risk-budgeted safe throughput forecasting problem.
  (a) Starlink throughput time-series: the predictor observes a historical window $X_t$ of length $L$ and forecasts the future sequence $Y_t$ over horizon $H$.
  (b) Conventional point prediction minimizes symmetric error (e.g., MAE) but treats overestimation and underestimation equally.
  (c) In the proposed formulation, overestimation ($\hat{y} > y$) is explicitly treated as a safety risk. The predictor must satisfy a prescribed overestimation budget $\varepsilon$ while maximizing accuracy. Underestimation is tolerated as a conservative but safe decision.}
  \label{fig:problem-overview}
\end{figure*}

This section formalizes safe throughput forecasting for LEO satellite networks. We first define the standard short-term forecasting task and then show why an accuracy-only objective is insufficient when overestimation can trigger harmful resource-allocation decisions. Based on this asymmetry, we formulate forecasting as a risk-budgeted constrained problem that controls overestimation while preserving accuracy. We further define positive-error severity metrics to capture not only how often a predictor overestimates, but also how large these overestimations are. Figure~\ref{fig:problem-overview} illustrates the problem formulation.

\subsection{Short-Term Throughput Forecasting}
We consider short-term throughput forecasting for Starlink-like LEO satellite access networks. Let $y_t$ denote the measured downlink throughput at time slot $t$. Given a historical observation window of length $L$, the input sequence is defined as
\begin{equation}
    X_t = [x_{t-L+1}, x_{t-L+2}, \ldots, x_t],
\end{equation}
where $x_t$ may include the historical throughput and other available network-side or measurement-side features. The forecasting target is the future throughput sequence over a prediction horizon of length $H$,
\begin{equation}
    Y_t = [y_{t+1}, y_{t+2}, \ldots, y_{t+H}].
\end{equation}
A conventional point forecasting model learns a mapping
\begin{equation}
    \hat{Y}_t = f_{\theta}(X_t),
\end{equation}
where $\theta$ denotes the model parameters. Such models are usually optimized by symmetric accuracy objectives, such as MAE or MSE. For example, the MAE objective can be written as
\begin{equation}
    A(\hat{Y},Y)=
    \frac{1}{NH}
    \sum_{t=1}^{N}
    \sum_{h=1}^{H}
    |\hat{y}_{t,h}-y_{t,h}|,
    \label{eq:mae}
\end{equation}
where $N$ is the number of evaluated samples, $y_{t,h}=y_{t+h}$ is the ground-truth throughput at horizon step $h$, and $\hat{y}_{t,h}$ is the corresponding predicted throughput. Here, $A(\hat{Y},Y)$ denotes the average absolute prediction error over all samples and horizon steps.

Although point forecasting is useful for characterizing the average prediction error, it does not distinguish between underestimation and overestimation. In network resource allocation, these two error types have different consequences. Underestimation mainly reduces resource utilization, whereas overestimation may lead to over-admission, bandwidth overbooking, queue buildup, packet loss, or service interruption. Therefore, an accuracy-only objective is insufficient for downstream decision modules that require safe bandwidth estimates.

\subsection{Risk-Budgeted Safe Throughput Forecasting}
To capture the asymmetric consequence of prediction errors, we formulate safe throughput forecasting as a risk-budget constrained problem. An overestimation event occurs when the predicted throughput exceeds the actual available throughput, i.e., $\hat{y}_{t,h}>y_{t,h}$. We define the overestimation rate as
\begin{equation}
    R(\hat{Y},Y)=
    \frac{1}{NH}
    \sum_{t=1}^{N}
    \sum_{h=1}^{H}
    \mathbb{I}(\hat{y}_{t,h}>y_{t,h}),
    \label{eq:overrate}
\end{equation}
where $\mathbb{I}(\cdot)$ is the indicator function, which equals one when its condition is true and zero otherwise. The application can specify a maximum acceptable overestimation budget $\varepsilon$. The goal of safe throughput forecasting is then
\begin{equation}
\begin{aligned}
    \min_{\hat{Y}^{safe}} \quad & A(\hat{Y}^{safe},Y) \\
    \text{s.t.} \quad & R(\hat{Y}^{safe},Y)\leq \varepsilon .
\end{aligned}
\label{eq:safe-objective}
\end{equation}
Here, $\hat{Y}^{safe}$ denotes the safe throughput forecast. This formulation does not require the safest possible prediction. Instead, it seeks the most accurate prediction among those satisfying the overestimation risk budget. This constrained view avoids driving the predictor toward an overly conservative solution that can waste usable satellite bandwidth and degrade application utility.

\subsection{Safety-Oriented Error Severity}
The overestimation rate in Equation~\ref{eq:overrate} measures how often a model overestimates the available throughput, but it does not quantify the severity of these overestimations. Therefore, we also use positive-error severity metrics. The mean positive error (MPE) is defined as
\begin{equation}
    \text{MPE}(\hat{Y},Y)=
    \frac{1}{NH}
    \sum_{t=1}^{N}
    \sum_{h=1}^{H}
    \max(\hat{y}_{t,h}-y_{t,h},0).
    \label{eq:mpe}
\end{equation}
We further use the 95th percentile of the positive-error component,
\begin{equation}
    \text{P95+Err} =
    P_{95}\left(\max(\hat{y}_{t,h}-y_{t,h},0)\right),
    \label{eq:p95pos}
\end{equation}
to characterize tail-risk overestimation, where $P_{95}(\cdot)$ denotes the empirical 95th percentile computed over all samples and horizon steps. These metrics complement MAE and RMSE by revealing whether a predictor produces severe positive deviations that may lead to harmful over-allocation in low-throughput periods.

\section{Budget-Guided Coarse-to-Fine Quantile Selection}
\begin{figure}[!t]
  \centering
  \includegraphics[width=\linewidth]{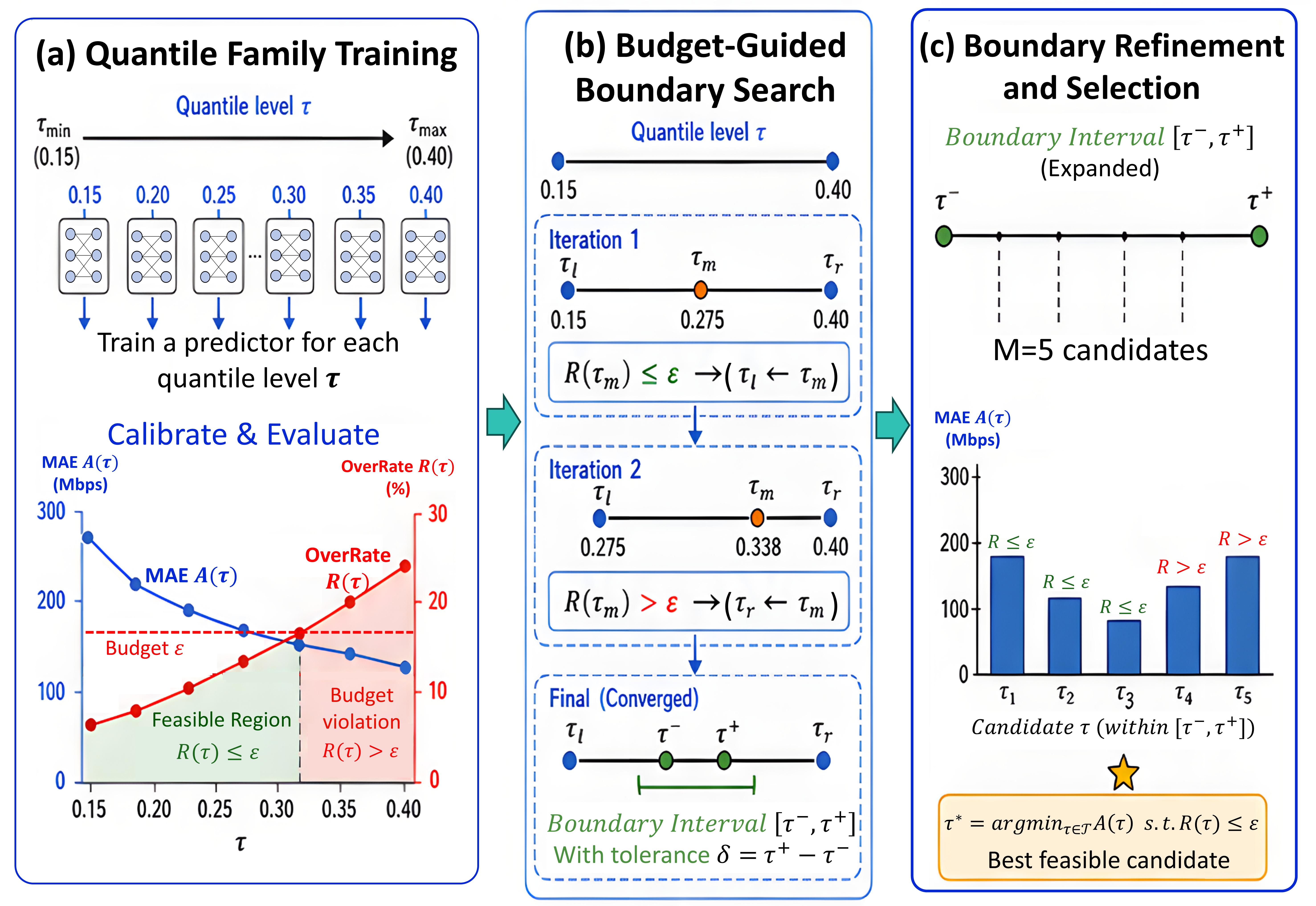}
  \caption{Overview of the BG-CFQS three-stage pipeline.
  \textit{Stage 1} --- Train a family of lower-quantile predictors over the candidate set $\mathcal{T}$ using pinball loss. As $\tau$ increases, the predictor becomes less conservative: accuracy improves but overestimation risk rises.
  \textit{Stage 2} --- Budget-guided boundary search: given a risk budget $\varepsilon$, iteratively evaluate midpoints (binary search) on the calibration set to locate the safety boundary $\tau^{-}$, the largest quantile satisfying $R(\tau) \leq \varepsilon$.
  \textit{Stage 3} --- Boundary refinement: construct a fine-grained grid within $[\tau^{-},\tau^{+}]$, evaluate all candidates, and select $\tau^{*}$ that minimizes MAE subject to the risk budget. If no feasible candidate exists, a penalized fallback objective is used.}
  \label{fig:bgcfqs-framework}
\end{figure}

This section presents BG-CFQS, a risk-calibrated quantile-selection framework for solving the safe forecasting objective in Equation~\ref{eq:safe-objective}. The method uses lower-quantile predictors as candidate safe forecasts and selects their operating quantile according to the prescribed overestimation budget. We derive the corresponding quantile-selection objective and describe the three stages of BG-CFQS illustrated in Figure~\ref{fig:bgcfqs-framework}.

\subsection{From Safe Forecasting to Quantile Selection}
The objective in Equation~\ref{eq:safe-objective} is defined over possible safe forecasts $\hat{Y}^{safe}$. Directly optimizing this objective is difficult because it does not specify how the forecast should be parameterized. BG-CFQS addresses this issue by restricting the candidate forecasts to a lower-quantile prediction family.

Let $\tau$ denote a quantile level. For a future throughput variable conditioned on the historical observation window, the $\tau$-quantile represents a lower-tail estimate of the future throughput distribution. In an ideal conditional quantile predictor, the true throughput falls below the predicted value with probability approximately $\tau$. Therefore, a smaller $\tau$ produces a more conservative throughput estimate, while a larger $\tau$ produces a less conservative estimate that may improve accuracy but increase overestimation risk.

Under this quantile-family parameterization, the safe forecast is written as
\begin{equation}
    \hat{Y}^{safe}_t = \hat{Y}^{(\tau^{*})}_t
    = f_{\theta}^{(\tau^{*})}(X_t),
\end{equation}
where $f_{\theta}^{(\tau)}$ is the predictor associated with quantile level $\tau$. For each candidate $\tau$, we evaluate its accuracy and risk on a calibration set:
\begin{equation}
    A(\tau)=A(\hat{Y}^{(\tau)},Y),
    \quad
    R(\tau)=R(\hat{Y}^{(\tau)},Y).
\end{equation}
Here, $A(\tau)$ and $R(\tau)$ are computed according to Equations~\ref{eq:mae} and~\ref{eq:overrate}, respectively. Let $\mathcal{T}$ denote the candidate set of lower-quantile levels considered by the method. The original safe forecasting problem can then be converted into a one-dimensional constrained selection problem:
\begin{equation}
    \tau^{*}
    =
    \arg\min_{\tau\in\mathcal{T}} A(\tau)
    \quad
    \text{s.t.}
    \quad
    R(\tau)\leq \varepsilon .
    \label{eq:tau-objective}
\end{equation}
Thus, $\tau$ is a risk-control variable selected from data under the prescribed overestimation budget. Based on this selection objective, BG-CFQS implements an efficient coarse-to-fine procedure instead of exhaustively evaluating all candidates in $\mathcal{T}$. The following subsections describe its three stages: quantile-family construction, budget-guided boundary search, and boundary refinement.

\subsection{Risk-Aware Quantile Family}
The first step of BG-CFQS is to construct the candidate lower-quantile family over which the risk budget will be enforced.

For a quantile level $\tau$, we train a lower-quantile predictor
\begin{equation}
    \hat{Y}^{(\tau)}_t = f_{\theta}^{(\tau)}(X_t).
\end{equation}
Here, $f_{\theta}^{(\tau)}$ denotes the predictor trained for quantile level $\tau$, and $\hat{Y}^{(\tau)}_t$ is its multi-step throughput forecast. A smaller $\tau$ usually produces a more conservative prediction, while a larger $\tau$ tends to improve accuracy but increases the probability of overestimation. For each $\tau$, the predictor is trained using the pinball loss
\begin{equation}
    \mathcal{L}_{\tau}(y,\hat{y})=
    \max\left(\tau(y-\hat{y}),(\tau-1)(y-\hat{y})\right).
    \label{eq:pinball}
\end{equation}
where $y$ and $\hat{y}$ denote a scalar ground-truth throughput and its prediction, respectively. For multi-step forecasting, the loss is averaged over the prediction horizon:
\begin{equation}
    \mathcal{L}_{\tau}(Y_t,\hat{Y}_t)=
    \frac{1}{H}
    \sum_{h=1}^{H}
    \mathcal{L}_{\tau}(y_{t,h},\hat{y}_{t,h}).
\end{equation}

The predictor family is trained for candidate lower-quantile levels in $\mathcal{T}$. In implementation, $\mathcal{T}$ is instantiated as a lower-quantile search interval below the median, with the specific range reported in the experimental setting.

\subsection{Budget-Guided Boundary Search}
Given the quantile-selection objective in Equation~\ref{eq:tau-objective}, BG-CFQS uses the calibration set to identify where the quantile family transitions from safe to unsafe. This step reduces the search cost by exploiting the empirical ordering of lower-quantile predictors instead of evaluating a dense grid over the entire search interval. For boundary search, we instantiate $\mathcal{T}$ as an interval $[\tau_{\min},\tau_{\max}]$.

For lower-quantile prediction, increasing $\tau$ typically makes the prediction less conservative. Therefore, $R(\tau)$ often exhibits an approximately increasing trend with respect to $\tau$. BG-CFQS exploits this empirical structure to search for the safety boundary instead of exhaustively evaluating all possible quantiles. When feasible candidates exist, the desired boundary is the largest quantile that still satisfies the budget:
\begin{equation}
    \tau^{-}=
    \max\{\tau\in[\tau_{\min},\tau_{\max}]:R(\tau)\leq\varepsilon\}.
    \label{eq:safe-boundary}
\end{equation}

The boundary search starts from $\tau_l=\tau_{\min}$ and $\tau_r=\tau_{\max}$. BG-CFQS first evaluates $R(\tau_l)$ and $R(\tau_r)$. If $R(\tau_r)\leq\varepsilon$, the whole interval is safe and the method sets the boundary to $\tau_r$. If $R(\tau_l)>\varepsilon$, even the most conservative endpoint violates the budget, and the method enters a budget-violation regime. Otherwise, the boundary lies between $\tau_l$ and $\tau_r$. In this case, BG-CFQS iteratively evaluates the midpoint
\begin{equation}
    \tau_m = \frac{\tau_l+\tau_r}{2}.
\end{equation}
If $R(\tau_m)\leq\varepsilon$, the safe side is extended by setting $\tau_l\leftarrow\tau_m$; otherwise, the unsafe side is updated by setting $\tau_r\leftarrow\tau_m$. The process stops when
\begin{equation}
    \tau_r-\tau_l < \delta,
\end{equation}
where $\delta$ is the coarse search tolerance. The search then returns a boundary interval $[\tau^{-},\tau^{+}]=[\tau_l,\tau_r]$, where $\tau^{-}$ is the safe-side boundary and $\tau^{+}$ is the adjacent boundary on the other side.

\subsection{Boundary Refinement and Final Selection}
The boundary search provides an efficient but coarse localization of the feasible frontier. BG-CFQS then refines this frontier to improve accuracy without using a hand-crafted local candidate set. Specifically, it constructs a fine-grained grid within the boundary interval:
\begin{equation}
    Q_{\text{fine}}=
    \text{LinSpace}(\tau^{-},\tau^{+},M),
    \label{eq:fine-grid}
\end{equation}
where $\text{LinSpace}(\tau^{-},\tau^{+},M)$ denotes $M$ evenly spaced quantile candidates between $\tau^{-}$ and $\tau^{+}$, including the two endpoints. Each candidate in $Q_{\text{fine}}$ is trained on the training set and evaluated on the calibration set.

The final quantile is selected by solving the constrained objective over the fine grid:
\begin{equation}
    \tau^{*}
    =
    \arg\min_{\tau\in Q_{\text{fine}}}
    A(\tau)
    \quad
    \text{s.t.}
    \quad
    R(\tau)\leq\varepsilon .
    \label{eq:fine-selection}
\end{equation}
If the feasible set is non-empty, BG-CFQS selects the candidate with the smallest MAE among all candidates satisfying the risk budget. If all fine-grid candidates violate the budget, BG-CFQS uses a penalized fallback objective:
\begin{equation}
    J(\tau)=
    A(\tau)+
    \lambda\cdot\max(R(\tau)-\varepsilon,0),
    \label{eq:penalty}
\end{equation}
and selects
\begin{equation}
    \tau^{*}
    =
    \arg\min_{\tau\in Q_{\text{fine}}}
    J(\tau),
\end{equation}
where $\lambda$ controls the penalty for violating the overestimation budget. The final safe throughput forecast is
\begin{equation}
    \hat{Y}^{safe}_t
    =
    f_{\theta}^{(\tau^{*})}(X_t).
    \label{eq:final-safe}
\end{equation}

Algorithm~\ref{alg:bgcfqs} summarizes BG-CFQS. The calibration set is used only for selecting the risk-calibrated quantile. After $\tau^{*}$ is determined, the selected model is evaluated on the test set using both accuracy and safety metrics.

\begin{algorithm}[!t]
\caption{Budget-Guided Coarse-to-Fine Quantile Selection}
\label{alg:bgcfqs}
\begin{algorithmic}[1]
\STATE \textbf{Input:} training set $\mathcal{D}_{train}$, calibration set $\mathcal{D}_{cal}$, quantile interval $[\tau_{\min},\tau_{\max}]$, risk budget $\varepsilon$, coarse tolerance $\delta$, fine grid size $M$, penalty weight $\lambda$
\STATE Train quantile predictors at $\tau_{\min}$ and $\tau_{\max}$ on $\mathcal{D}_{train}$
\STATE Evaluate $R(\tau_{\min})$ and $R(\tau_{\max})$ on $\mathcal{D}_{cal}$
\IF{$R(\tau_{\max})\leq\varepsilon$}
    \STATE Set boundary interval $[\tau^{-},\tau^{+}]=[\tau_{\max},\tau_{\max}]$
\ELSIF{$R(\tau_{\min})>\varepsilon$}
    \STATE Set boundary interval $[\tau^{-},\tau^{+}]=[\tau_{\min},\tau_{\min}]$
\ELSE
    \STATE Set $\tau_l=\tau_{\min}$ and $\tau_r=\tau_{\max}$
    \WHILE{$\tau_r-\tau_l\geq\delta$}
        \STATE $\tau_m \leftarrow (\tau_l+\tau_r)/2$
        \STATE Train $f_{\theta}^{(\tau_m)}$ on $\mathcal{D}_{train}$ and evaluate it on $\mathcal{D}_{cal}$
        \IF{$R(\tau_m)\leq\varepsilon$}
            \STATE $\tau_l \leftarrow \tau_m$
        \ELSE
            \STATE $\tau_r \leftarrow \tau_m$
        \ENDIF
    \ENDWHILE
    \STATE Set boundary interval $[\tau^{-},\tau^{+}]=[\tau_l,\tau_r]$
\ENDIF
\STATE Build $Q_{\text{fine}}=\text{LinSpace}(\tau^{-},\tau^{+},M)$
\STATE Train all candidates in $Q_{\text{fine}}$ on $\mathcal{D}_{train}$ and evaluate them on $\mathcal{D}_{cal}$
\IF{$\exists \tau\in Q_{\text{fine}}$ such that $R(\tau)\leq\varepsilon$}
    \STATE Select $\tau^{*}=\arg\min_{\tau\in Q_{\text{fine}}, R(\tau)\leq\varepsilon} A(\tau)$
\ELSE
    \STATE Select $\tau^{*}=\arg\min_{\tau\in Q_{\text{fine}}} A(\tau)+\lambda\max(R(\tau)-\varepsilon,0)$
\ENDIF
\STATE \textbf{Output:} selected quantile $\tau^{*}$ and safe predictor $f_{\theta}^{(\tau^{*})}$
\end{algorithmic}
\end{algorithm}

\subsection{Application-Level Interpretation}
The proposed formulation is intended to support downstream network-control modules that require a single conservative bandwidth estimate. We illustrate this interpretation with an admission-control example. Suppose each incoming service requires $b$ Mbps. Given a predicted safe throughput $\hat{y}^{safe}$, the number of admitted services is
\begin{equation}
    n_{\text{admit}}=
    \left\lfloor
    \frac{\hat{y}^{safe}}{b}
    \right\rfloor .
\end{equation}
The oracle number of services that can be supported by the actual throughput $y$ is
\begin{equation}
    n_{\text{oracle}}=
    \left\lfloor
    \frac{y}{b}
    \right\rfloor .
\end{equation}
The number of successfully served services and dropped services are then
\begin{equation}
    n_{\text{served}}=
    \min(n_{\text{admit}},n_{\text{oracle}}),
\end{equation}
\begin{equation}
    n_{\text{drop}}=
    \max(n_{\text{admit}}-n_{\text{oracle}},0).
\end{equation}
This decision-level interpretation connects forecasting safety to application consequences. A model that frequently overestimates throughput may admit too many services and increase $n_{\text{drop}}$, even if its MAE is low. By explicitly controlling overestimation risk, BG-CFQS aims to reduce such harmful decisions while maintaining useful bandwidth utilization.

\section{Experiment}
\subsection{Experimental Setting}
\label{sec:experimental-setting}
\subsubsection{Datasets and Prediction Task}
We evaluate BG-CFQS on three Starlink throughput datasets~\cite{liu2025starnet} derived from the StarNet measurement study. The datasets are collected from three geographical regions and are denoted as \emph{CHI}, \emph{OSN}, and \emph{VIC} in our experiments. Table~\ref{tab:dataset-summary} summarizes both the original measurement statistics reported by StarNet~\cite{liu2025starnet} and the processed data files used in our experiments. The dataset contains three country-level traces from the United States, Germany, and Canada. In our experiments, CHI corresponds to the processed U.S. trace, OSN corresponds to the processed Germany trace, and VIC corresponds to the processed Canada/Victoria trace. 

\begin{table*}[!t]
\centering
\caption{Summary of Starlink throughput datasets.}
\label{tab:dataset-summary}
\begin{tabular}{lccc}
\toprule
\textbf{Item} & \textbf{CHI} & \textbf{OSN} & \textbf{VIC} \\
\midrule
Location & Chicago, USA & OSN, Germany & Victoria, Canada \\
StarNet split & USA & Germany & Canada \\
Dishy version & rev3\_proto2 & rev3\_proto2 & rev3\_proto2 \\
Original duration & 6 months & 1 month & 1 month \\
Cumulative trace minutes & 41,252 & 10,221 & 2,417 \\
Original throughput samples & 2,475,163 & 613,295 & 145,053 \\
Unique serving satellites & 6,052 & 3,956 & 3,166 \\
Satellite handovers & 86,808 & 26,782 & 7,257 \\
Processed time range & 2024-04-26--2024-05-28 & 2024-07-13--2024-07-31 & 2024-07-11--2024-07-28 \\
Used samples & 1,123,832 & 613,295 & 145,053 \\
\bottomrule
\end{tabular}
\end{table*}

For each prediction sample, the model observes a fixed-length multivariate history window and predicts the future throughput sequence over the next $H$ slots. The input window contains historical throughput together with auxiliary variables available in the processed Starlink traces. These auxiliary variables include satellite geometry and association features, i.e., elevation angle, azimuth angle, satellite distance, encoded satellite identifier, and the number of candidate satellites; weather features, i.e., cloud coverage, pressure, and humidity; and time features derived from the timestamp, i.e., 15-second phase, minute, hour, and day of week. The raw timestamp and latency are excluded from the model inputs. XGBoost-based predictors use a flattened representation of this history window, whereas sequence models consume the same window in sequence form. The same prediction setting is used for all compared methods, and all safety metrics are computed over all prediction horizons unless otherwise specified. Table~\ref{tab:exp-setting} summarizes the main experimental configuration.

\begin{table}[!t]
\centering
\caption{Experimental configuration.}
\label{tab:exp-setting}
\begin{tabular}{lc}
\toprule
\textbf{Item} & \textbf{Value} \\
\midrule
History length $L$ & 75 \\
Prediction horizon $H$ & 15 \\
Default risk budget $\varepsilon$ & 0.35 \\
Candidate quantile set $\mathcal{T}$ & $[0.15,0.40]$ \\
Coarse tolerance $\delta$ & 0.05 \\
Fine-grid size $M$ & 5 \\
\bottomrule
\end{tabular}
\end{table}

\subsubsection{Baselines}
We compare BG-CFQS with both point forecasting baselines and risk-calibrated baselines. The point forecasting baselines include T3P-point, StarNet-point, and DLinear-point. Following T3P~\cite{tiwari2023t3p}, we implement the T3P baseline with its XGBoost-based throughput predictor~\cite{chen2016xgboost}, while StarNet denotes the satellite-domain sequence prediction model proposed in~\cite{liu2025starnet}. These methods directly output point throughput predictions and are mainly optimized for conventional prediction accuracy. In the proposed BG-CFQS implementation, each lower-quantile predictor also uses an XGBoost backbone~\cite{chen2016xgboost} trained with the pinball loss at the corresponding quantile level. 

To ensure a fair comparison under the same risk budget, we further construct budget-scale variants for each point predictor. Given a point prediction $\hat{Y}$, the budget-scale baseline outputs
\begin{equation}
    \hat{Y}^{safe}=c\hat{Y},
\end{equation}
where the scaling factor $c$ is selected on the calibration set by
\begin{equation}
\begin{aligned}
    c^{*}=
    \arg\min_{c} \quad & A(c\hat{Y},Y)\\
    \text{s.t.} \quad & R(c\hat{Y},Y)\leq\varepsilon .
\end{aligned}
\end{equation}
Accordingly, the primary risk-calibrated baselines are T3P-budget-scale, StarNet-budget-scale, and DLinear-budget-scale. 

For the separate generality analysis, we additionally include PatchTST~\cite{nie2022time}, TimesNet~\cite{wu2022timesnet}, and iTransformer~\cite{liu2023itransformer}, together with their budget-scale variants calibrated by the same constrained scaling rule.

\subsubsection{Training, Calibration, and Testing Protocol}
For each dataset, we use a training/calibration/test protocol. The training set is used to fit the point predictors and quantile predictors. The calibration set is used only for risk-control selection, including the scaling factor of budget-scale baselines and the final quantile of BG-CFQS. The test set is held out for final reporting and is not used for model fitting or risk-control selection.

For point forecasting baselines, the raw model output is evaluated directly on the test set. For budget-scale baselines, the scaling factor is selected on the calibration set and then applied to the test set. For BG-CFQS, the boundary search and fine-grid selection are performed on the calibration set, and the selected quantile predictor is evaluated on the test set. This protocol ensures that risk calibration and final evaluation are separated.

\subsubsection{Evaluation Metrics}
We report conventional accuracy metrics, including MAE and RMSE. To evaluate safety, we report OverRate, MPE, and P95+Err, as defined in Section III. We also evaluate model behavior in low-throughput regimes. Specifically, \emph{High-risk P30} denotes samples whose true throughput belongs to the lowest 30\% of a dataset, and \emph{Severe-risk P10} denotes samples whose true throughput belongs to the lowest 10\%. These subsets represent safety-critical operating regimes in which throughput overestimation can directly amplify over-allocation risk.

\subsection{Overall Risk-Budgeted Forecasting Performance}
We evaluate whether accuracy-oriented throughput predictors satisfy the safety requirement needed by downstream network-control decisions. A predictor with low MAE or RMSE can still be unsuitable if it overestimates available throughput too frequently. Following the protocol in Section~\ref{sec:experimental-setting}, we compare raw point predictors, budget-scale variants, and BG-CFQS on the same test samples and prediction horizon. Table~\ref{tab:budget-average} reports the average results over CHI, OSN, and VIC; Risk Pass denotes the number of datasets satisfying the OverRate budget.

Point forecasting methods achieve relatively low MAE, but all of them violate the risk budget. For example, T3P-point obtains the lowest average MAE of 37.775, but its average OverRate is 0.512, which substantially exceeds the default budget. Similar behavior can also be observed for StarNet-point and DLinear-point. After risk calibration, all budget-scale baselines satisfy the risk budget, confirming that simple conservative calibration is a strong and necessary baseline. However, compared with these risk-calibrated baselines, BG-CFQS achieves the best average MAE among all budget-feasible methods and also obtains the lowest MPE and P95+Err. This indicates that BG-CFQS does not merely reduce the frequency of overestimation; it also reduces the magnitude and tail severity of harmful overestimation.

\begin{table*}[!t]
\centering
\caption{Average risk-budgeted forecasting performance.}
\label{tab:budget-average}
\begin{tabular}{lcccccc}
\toprule
\textbf{Method} & \textbf{MAE} & \textbf{RMSE} & \textbf{OverRate} & \textbf{MPE} & \textbf{P95+Err} & \textbf{Risk Pass} \\
\midrule
T3P-point & 37.775 & 49.699 & 0.512 & 20.182 & 87.984 & 0/3 \\
T3P-budget-scale & 40.985 & 52.305 & 0.346 & 12.444 & 69.553 & 3/3 \\
StarNet-point & 39.806 & 52.102 & 0.433 & 17.030 & 82.842 & 0/3 \\
StarNet-budget-scale & 42.154 & 54.061 & 0.348 & 13.027 & 72.691 & 3/3 \\
DLinear-point & 42.192 & 54.315 & 0.500 & 21.791 & 93.479 & 0/3 \\
DLinear-budget-scale & 45.787 & 57.869 & 0.348 & 13.372 & 73.533 & 3/3 \\
BG-CFQS & \textbf{40.364} & 52.480 & 0.349 & \textbf{11.745} & \textbf{65.834} & 3/3 \\
\bottomrule
\end{tabular}
\end{table*}

\subsection{Dataset-Level Risk Calibration}
We examine whether risk calibration should be dataset-dependent. Starlink throughput dynamics vary across locations because of satellite visibility, traffic load, weather, and deployment conditions; thus, a single conservative level may not provide the best safety-accuracy tradeoff for all traces. Table~\ref{tab:dataset-risk-budget} reports only budget-feasible methods on each dataset. For BG-CFQS, the Control column denotes the selected quantile $\tau^{*}$; for budget-scale baselines, it denotes the selected scaling factor $c^{*}$. Boldface marks the best budget-feasible value for the key accuracy and positive-error metrics in each dataset.

The selected BG-CFQS quantile differs across datasets: 0.314 for CHI, 0.244 for OSN, and 0.306 for VIC. This supports the motivation that the safe prediction level should be selected according to data and risk budget, rather than manually fixed in advance. The scale factors selected by the budget-scale baselines also vary across datasets and backbones, indicating that the required degree of conservativeness is both data-dependent and model-dependent. Under comparable OverRate values, BG-CFQS achieves the lowest MPE and P95+Err on all three datasets, showing that it consistently reduces the severity of harmful overestimation. In terms of MAE, BG-CFQS performs best on OSN and VIC; on CHI, StarNet-budget-scale obtains a lower MAE, while BG-CFQS still provides the lowest positive-error severity. Therefore, the dataset-level results support both aspects of the proposed calibration: the selected quantile adapts to heterogeneous traces, and the resulting safe predictor improves overestimation severity without losing competitive accuracy.

\begin{table*}[!t]
\centering
\caption{Dataset-level comparison among budget-feasible methods.}
\label{tab:dataset-risk-budget}
\begin{tabular}{llccccc}
\toprule
\textbf{Dataset} & \textbf{Method} & \textbf{Control} & \textbf{MAE} & \textbf{OverRate} & \textbf{MPE} & \textbf{P95+Err} \\
\midrule
CHI & T3P-budget-scale & 0.920 & 44.324 & 0.348 & 14.657 & 81.238 \\
CHI & StarNet-budget-scale & 0.920 & \textbf{43.191} & 0.349 & 14.552 & 80.794 \\
CHI & DLinear-budget-scale & 0.900 & 46.123 & 0.348 & 15.018 & 83.031 \\
CHI & BG-CFQS & 0.314 & 44.286 & 0.349 & \textbf{13.839} & \textbf{76.913} \\
\midrule
OSN & T3P-budget-scale & 0.870 & 39.091 & 0.348 & 12.432 & 70.206 \\
OSN & StarNet-budget-scale & 0.970 & 38.626 & 0.347 & 12.124 & 68.804 \\
OSN & DLinear-budget-scale & 0.930 & 42.055 & 0.348 & 12.619 & 69.980 \\
OSN & BG-CFQS & 0.244 & \textbf{38.014} & 0.349 & \textbf{11.553} & \textbf{65.722} \\
\midrule
VIC & T3P-budget-scale & 0.890 & 39.541 & 0.342 & 10.243 & 57.217 \\
VIC & StarNet-budget-scale & 0.940 & 44.646 & 0.347 & 12.403 & 68.476 \\
VIC & DLinear-budget-scale & 0.840 & 49.182 & 0.348 & 12.479 & 67.587 \\
VIC & BG-CFQS & 0.306 & \textbf{38.792} & 0.347 & \textbf{9.844} & \textbf{54.868} \\
\bottomrule
\end{tabular}
\end{table*}

\subsection{Generality Across Advanced Time-Series Forecasting Models}
We test whether the safety issue persists for stronger general-purpose time-series forecasting architectures. This experiment compares BG-CFQS with PatchTST~\cite{nie2022time}, TimesNet~\cite{wu2022timesnet}, and iTransformer~\cite{liu2023itransformer}, using the same train/calibration/test protocol and prediction setting. For each backbone, we evaluate both the raw point predictor and a budget-scale variant calibrated by the same constrained scaling rule used in Section~\ref{sec:experimental-setting}. Table~\ref{tab:advanced-ts} reports the average results over CHI, OSN, and VIC, with boldface marking the best budget-feasible value for the key accuracy and positive-error metrics.

The results show that advanced point forecasting models are still not inherently safe. PatchTST-point, TimesNet-point, and iTransformer-point obtain average OverRate values of 0.444, 0.463, and 0.703, respectively, and therefore fail the risk budget on most datasets. After budget-scale calibration, all three advanced backbones satisfy the budget on all datasets, but their positive-error severity remains higher than BG-CFQS. Compared with the strongest advanced budget-scale baseline, PatchTST-budget-scale, BG-CFQS reduces MAE from 44.695 to 40.364, MPE from 13.299 to 11.745, and P95+Err from 73.378 to 65.834. This confirms that the benefit of BG-CFQS is not limited to a small set of baselines: directly selecting a lower-quantile predictor under a risk budget provides a stronger safety-accuracy tradeoff than post-hoc scaling even for modern time-series forecasting architectures.

\begin{table*}[!t]
\centering
\caption{Average comparison with advanced time-series forecasting models.}
\label{tab:advanced-ts}
\begin{tabular}{lcccccc}
\toprule
\textbf{Method} & \textbf{MAE} & \textbf{RMSE} & \textbf{OverRate} & \textbf{MPE} & \textbf{P95+Err} & \textbf{Risk Pass} \\
\midrule
PatchTST-point & 43.306 & 55.928 & 0.444 & 18.529 & 85.412 & 1/3 \\
PatchTST-budget-scale & 44.695 & 56.667 & 0.345 & 13.299 & 73.378 & 3/3 \\
TimesNet-point & 42.569 & 54.989 & 0.463 & 19.843 & 89.458 & 0/3 \\
TimesNet-budget-scale & 45.206 & 57.552 & 0.348 & 13.394 & 73.732 & 3/3 \\
iTransformer-point & 64.220 & 80.075 & 0.703 & 50.711 & 153.862 & 0/3 \\
iTransformer-budget-scale & 59.730 & 74.788 & 0.348 & 16.037 & 86.342 & 3/3 \\
BG-CFQS & \textbf{40.364} & 52.480 & 0.349 & \textbf{11.745} & \textbf{65.834} & 3/3 \\
\bottomrule
\end{tabular}
\end{table*}

\subsection{Risk-Accuracy Frontier}
\begin{figure*}[!t]
  \centering
  \includegraphics[width=0.95\textwidth]{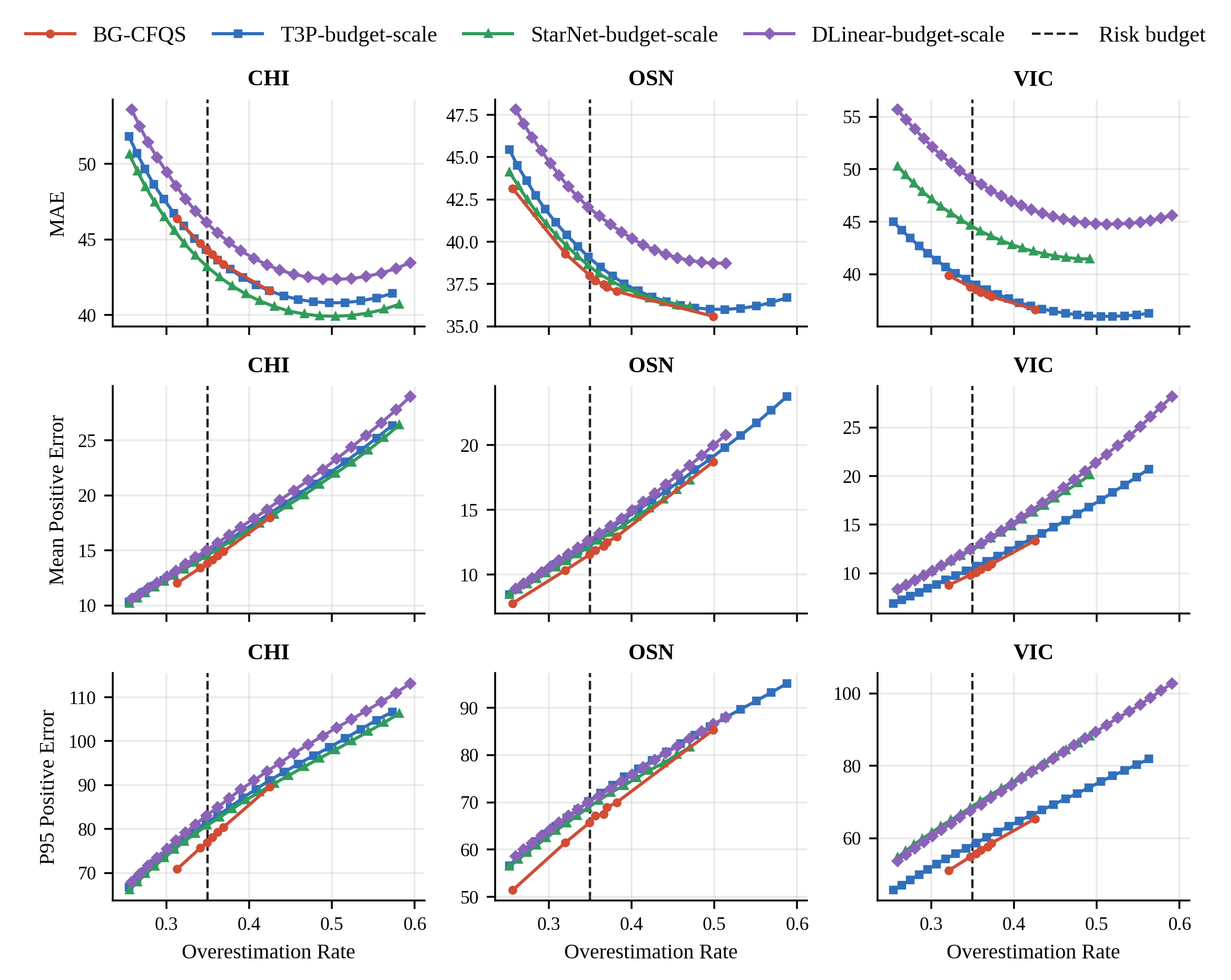}
\caption{Risk-accuracy frontiers across risk-control operating points.}
  \label{fig:risk-frontier}
\end{figure*}

A single risk budget gives only one operating point, whereas different network applications may require different levels of conservativeness. We therefore evaluate the full risk-accuracy frontier by sweeping the target overestimation budget from 0.30 to 0.50. For each budget, BG-CFQS re-runs the budget-guided selection procedure on the calibration set, while each scale-based baseline re-selects its multiplicative factor using the same constrained objective. Figure~\ref{fig:risk-frontier} plots MAE, MPE, and P95+Err against the achieved OverRate. Each point corresponds to one calibrated operating point; curves closer to the lower-left region indicate a better tradeoff, and at a similar OverRate the lower curve is preferable. The dashed vertical line marks the default operating budget.

Consistent with the risk-accuracy tradeoff, relaxing the risk budget generally improves MAE but increases overestimation severity. In the MAE row, BG-CFQS remains close to the best scale-based frontier on all datasets, indicating that the proposed quantile selection does not excessively sacrifice point accuracy. The advantage becomes clearer in the MPE and P95+Err rows: around the target risk budget and under comparable OverRate values, BG-CFQS is usually located below the scale-based baselines, especially on OSN and VIC. This means that BG-CFQS reduces the magnitude of harmful overestimation rather than merely matching the overestimation frequency. These results support the proposed risk-aware formulation, where the objective is not only to reduce average error but also to control the severity of positive prediction errors.

\subsection{High-Risk and Severe-Risk Analysis}
\begin{figure*}[!t]
  \centering
  \includegraphics[width=0.95\textwidth]{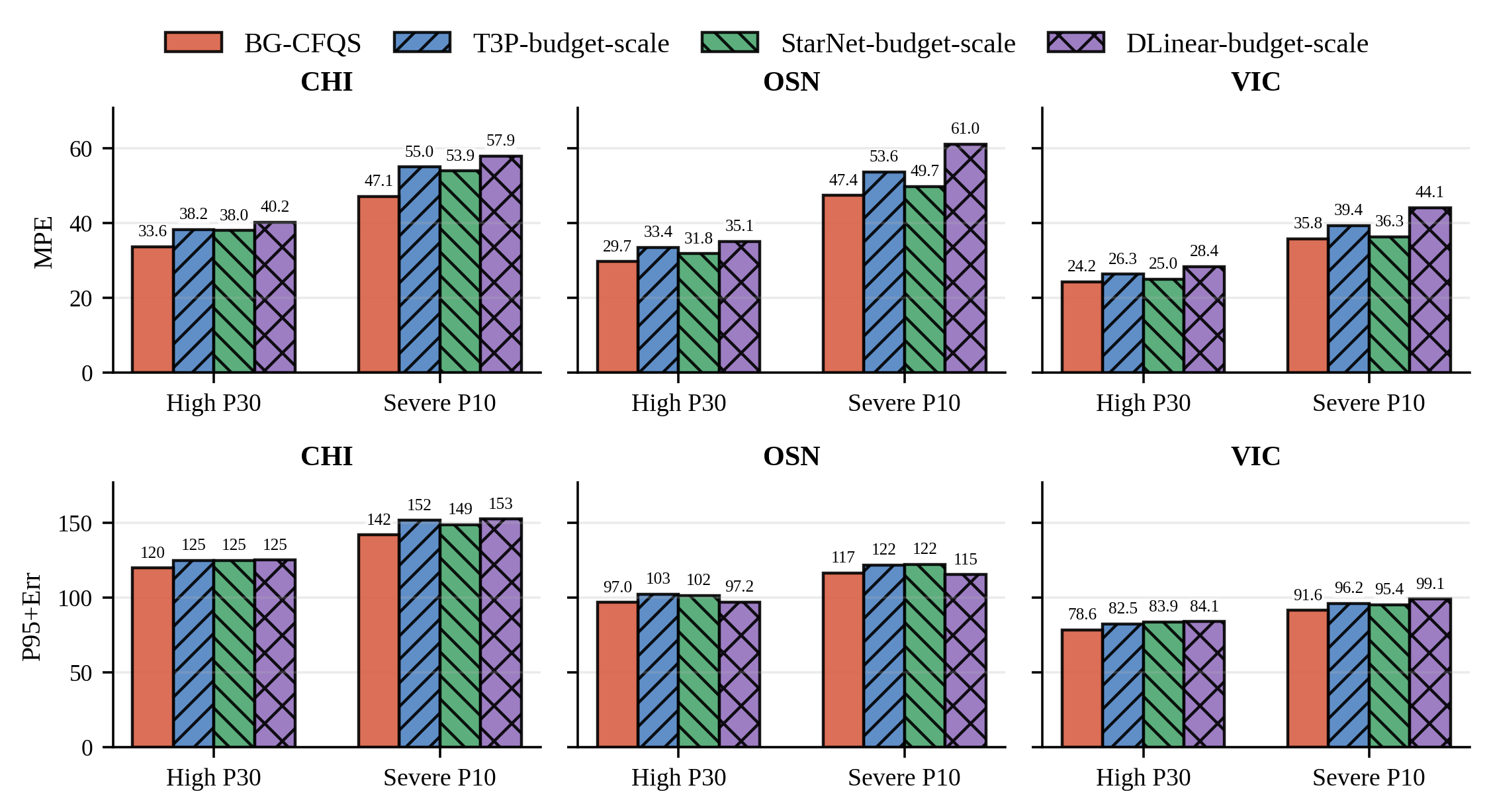}
\caption{Positive-error severity on low-throughput subsets.}
  \label{fig:high-risk-mpe-p95}
\end{figure*}

\begin{figure*}[!t]
  \centering
  \includegraphics[width=0.95\textwidth]{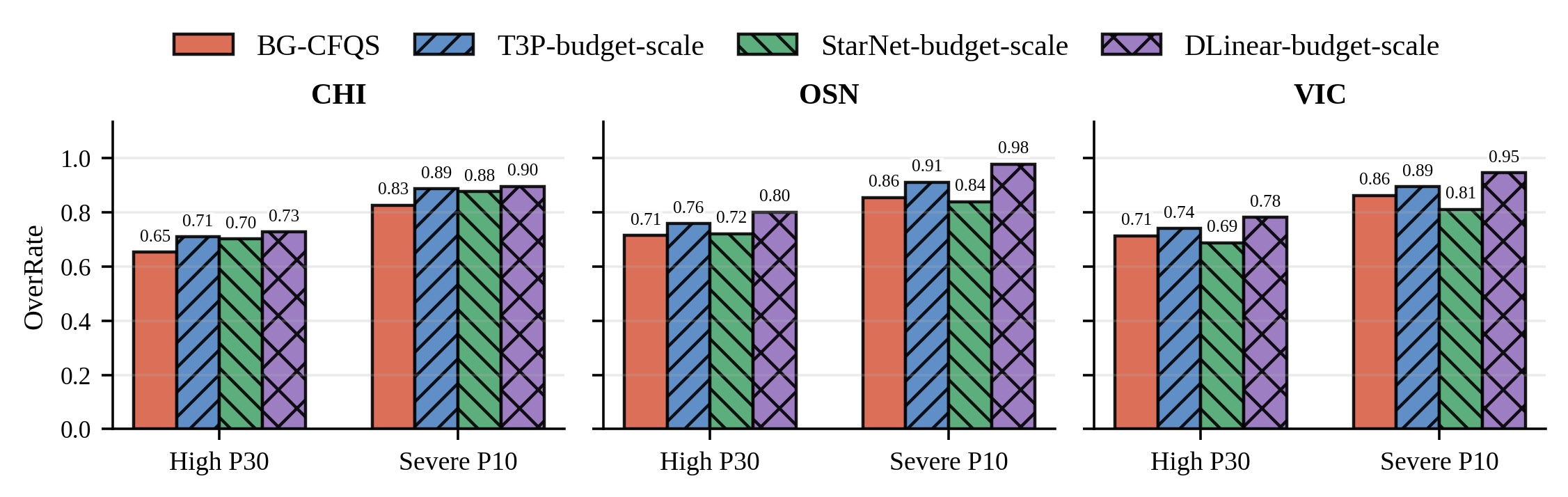}
\caption{OverRate on low-throughput subsets.}
  \label{fig:high-risk-overrate}
\end{figure*}

Global metrics can obscure unsafe behavior in low-capacity periods, where throughput overestimation is most likely to cause over-admission or bandwidth overbooking. We therefore evaluate safety on low-throughput subsets. For each dataset, test samples are ranked by true throughput; High-risk P30 contains the lowest 30\% of samples, and Severe-risk P10 contains the lowest 10\%. Within these subsets, we compute OverRate, MPE, and P95+Err. BG-CFQS is compared only with globally budget-feasible baselines, namely T3P-budget-scale, StarNet-budget-scale, and DLinear-budget-scale, on CHI, OSN, and VIC. Table~\ref{tab:high-risk-reduction} reports the average relative reduction over these baselines and datasets, while Figures~\ref{fig:high-risk-mpe-p95} and~\ref{fig:high-risk-overrate} provide detailed subset-level comparisons.

In the High-risk P30 subset, BG-CFQS reduces MPE by 11.0\% and P95+Err by 4.5\%. In the more challenging Severe-risk P10 subset, BG-CFQS reduces MPE by 12.6\% and P95+Err by 4.7\%. Figure~\ref{fig:high-risk-mpe-p95} shows that the reduction in positive-error severity is consistent across datasets: BG-CFQS achieves lower MPE than all three budget-scale baselines in both low-throughput subsets, and it also reduces P95+Err in nearly all pairwise comparisons. Figure~\ref{fig:high-risk-overrate} shows a different but complementary trend. When evaluation is restricted to low-throughput samples, OverRate increases for all methods because the true capacity is already near the lower tail of each dataset. Under this conditional evaluation, BG-CFQS keeps OverRate comparable to the budget-scale baselines, while producing smaller positive-error magnitudes. This indicates that the proposed method improves the severity of harmful overestimation in low-capacity regimes rather than merely shifting the operating point to an overly conservative predictor.

\begin{table}[!t]
\centering
\caption{Average reduction on low-throughput subsets.}
\label{tab:high-risk-reduction}
\begin{tabular}{lcc}
\toprule
\textbf{Subset} & \textbf{MPE Reduction} & \textbf{P95+Err Reduction} \\
\midrule
High-risk P30 & 11.0\% & 4.5\% \\
Severe-risk P10 & 12.6\% & 4.7\% \\
\bottomrule
\end{tabular}
\end{table}

\subsection{Admission-Control Evaluation}
\begin{figure*}[!t]
\centering
  \includegraphics[width=0.95\textwidth]{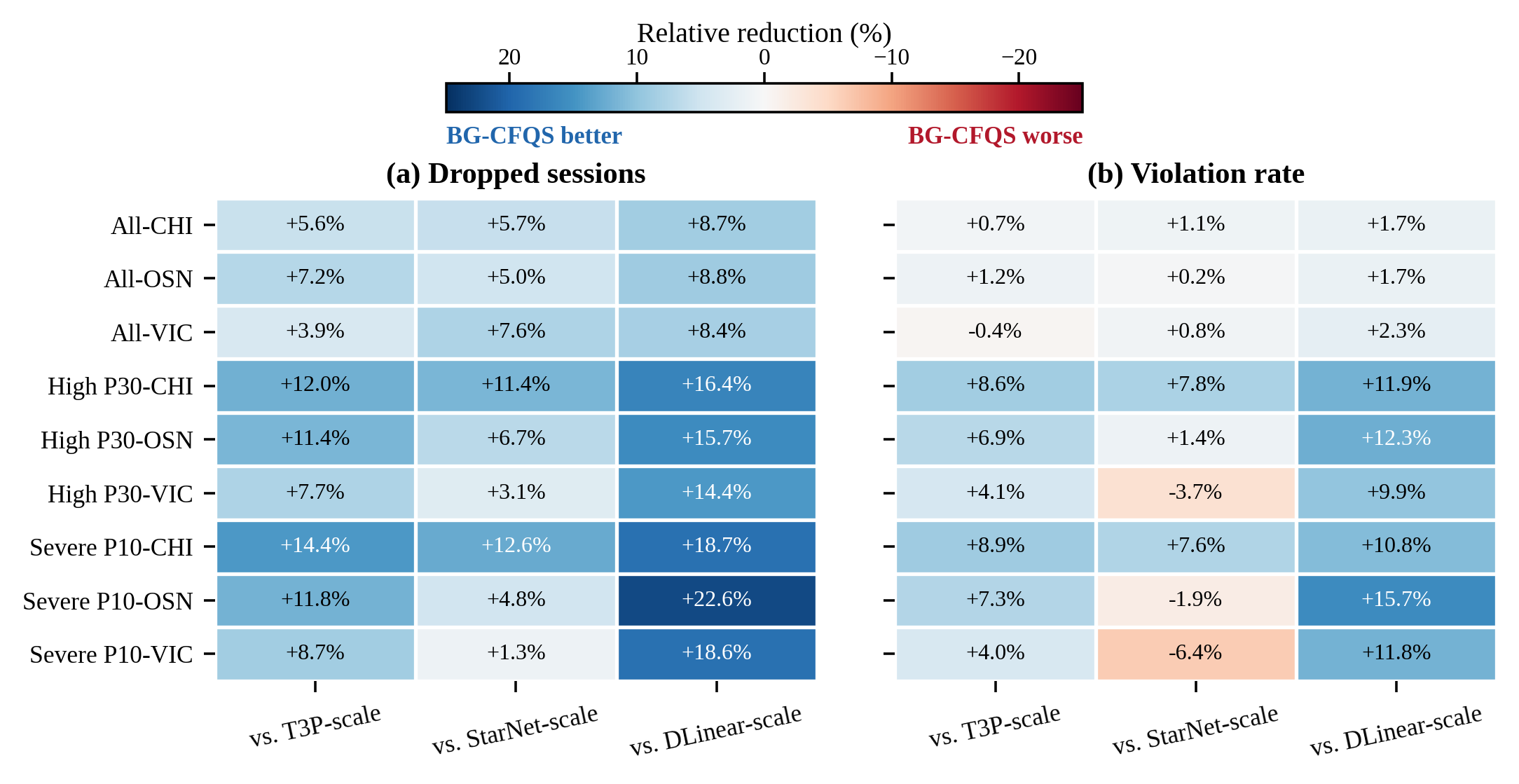}
\caption{Admission-control relative reduction of BG-CFQS over budget-scale baselines.}
  \label{fig:admission-control}
\end{figure*}

We evaluate whether safer forecasts lead to more reliable downstream decisions. Following the admission-control formulation in Section IV, we consider a slot-level setting in which each admitted service requires $b=10$ Mbps. For each prediction slot, a method admits $\lfloor \hat{y}^{safe}/b \rfloor$ sessions, while the oracle capacity is $\lfloor y/b \rfloor$; any excess admission is counted as dropped sessions. We report mean dropped sessions, violation rate, and P95 dropped sessions on all decisions, High-risk P30 samples, and Severe-risk P10 samples. BG-CFQS is compared with T3P-budget-scale, StarNet-budget-scale, and DLinear-budget-scale on CHI, OSN, and VIC. Table~\ref{tab:admission-reduction} reports the average relative reduction of BG-CFQS over these baselines and datasets, and Figure~\ref{fig:admission-control} further breaks down the relative reduction against each baseline. Positive blue values indicate fewer dropped sessions or violations, while negative red values indicate an increase.

As shown in Table~\ref{tab:admission-reduction}, BG-CFQS reduces the mean number of dropped sessions by 6.8\% on all decisions, while also slightly reducing the violation rate. The improvement becomes more pronounced in risk-sensitive regimes: BG-CFQS reduces dropped sessions by 11.0\% in High-risk P30 and 12.6\% in Severe-risk P10, with violation-rate reductions of 6.6\% and 6.4\%, respectively. Figure~\ref{fig:admission-control} further decomposes these averages by baseline and subset. Most cells are positive, indicating that BG-CFQS usually admits fewer excess sessions or incurs fewer violation events than the corresponding budget-scale baseline. The gains are concentrated in High-risk P30 and Severe-risk P10, where capacity is scarce and over-admission is more likely. These results confirm that the proposed safe forecasting framework does not only improve risk-aware prediction metrics, but can also reduce harmful admission-control decisions.

\begin{table}[!t]
\centering
\caption{Average relative reduction in admission-control decisions.}
\label{tab:admission-reduction}
\begin{tabular}{lcc}
\toprule
\textbf{Subset} & \textbf{Dropped Sessions} & \textbf{Violation Rate} \\
\midrule
All & 6.8\% & 1.0\% \\
High-risk P30 & 11.0\% & 6.6\% \\
Severe-risk P10 & 12.6\% & 6.4\% \\
\bottomrule
\end{tabular}
\end{table}

\section{Conclusion}
This paper presented a risk-aware safe forecasting framework for Starlink throughput prediction. Instead of treating throughput prediction as accuracy-oriented point forecasting, we reformulated it as a risk-budget constrained problem that explicitly controls overestimation before forecasts are used by resource-management applications. We then developed BG-CFQS to automatically identify a data-dependent lower-quantile operating point under a prescribed overestimation budget, and constructed a safety-oriented evaluation protocol covering overestimation frequency, positive-error severity, high-risk and severe-risk low-throughput regimes, and an admission-control application scenario. Across three real-world Starlink datasets, BG-CFQS satisfies the risk budget on all datasets and achieves the lowest average MAE, MPE, and P95+Err among budget-feasible methods. The selected quantile varies across datasets, confirming the need for data-driven risk calibration in heterogeneous LEO operating environments. Moreover, the improvements in low-throughput regimes and admission-control decisions show that safer forecasts can reduce harmful prediction errors where they matter most. Future work will extend this risk-budgeted view to sequential decision settings with time-coupled risk budgets and to other safety-critical network prediction targets such as latency and link quality.

\bibliographystyle{IEEEtran}
\bibliography{ref}

\vfill

\end{document}